# Boson Sampling Using Fock States from On-Demand Single Photon Source Arrays: Quantum Complexity & Impact of Dephasing

Swarnabha Chattaraj[1,2] and Anupam Madhukar[1]

[1] Nanostructure Materials and Devices Laboratory, University of Southern California

[2] Currently at Argonne National Laboratory

**Abstract:**

Boson sampling with array of on-demand single photons (discreet variable Fock states) as resource states provides distinct advantages in complexity scaling compared to single photon resource states mimicked by the states (continuous variable) generated in nonlinear photonics. Recent progress in realization of regular ordered array of mesa top single quantum dots (MTSQDs) enabling on-demand generation of many photon resource states with high single photon purity and indistinguishability [1,2] has opened the possibility of implementing chip-scale Boson sampling with single photons. An important question towards scale up of such circuits is how the dephasing and decay timescale of individual on-demand single photons affect the final probability distribution of the N-photon entangled state. In this work we report a framework that provides direct link between the physical sources of decoherence to the outcome of multiphoton correlation measurements. We explicitly incorporate the symmetrization of the N-photon states while accounting for the shape of the individual wave packets and dephasing. Using this framework, we show predictive estimates of the probability of generation of a 4-photon NOON state using an optical circuit initialized with 4 identical single photon wave packets, mimicking the emitted photons from MTSQD single photon sources reported in [2]. Specifically, we show how the probability of generating various 4-photon entangled states depend on the two-photon interference visibility which is determined by the pure dephasing timescale in relation to the radiative decay timescale. The results provide quantitative guidelines to measure 4 photon and higher number of entangled photon states generated in an optical circuit with on demand single photon sources and provide a link between physical processes such as temperature and the output state of the quantum optical circuits.

## I. Introduction

Information, however defined, is implicitly measured in discrete units (i.e. is quantized) and is stored and manipulated (processed) currently dominantly using the rules of classical physics but much effort is being expended-- and progress made-- on processing information using the rules of quantum mechanics. The dominant part of quantum information processing (QIP) on the public radar screen is quantum *computation* which, amongst a few paradigms, can be accomplished through the paradigm of measurement-based quantum computing (MBQC). The MBQC, in turn, can be implemented using two approaches dubbed "qubits" and "qumodes" with their own strengths and weaknesses originating in the two most basic features of quantum mechanics: (1) that all observable features are "quantized" (i.e. have discrete values) and (2) that each observable feature has a counterpart such that both (dubbed conjugate variables) cannot be measured simultaneously with arbitrary precision—a feature embodied in Heisenberg's uncertainty principle given in 1927.

In 1981, Caves showed [3] that within the Heisenberg principle, one can increase the precision of one variable (in principle arbitrarily) by sacrificing that (i.e. losing knowledge) of the other. This has been dubbed "squeezing" one variable at the expense of its conjugate. In such a situation, for instance for $\Delta X \rightarrow 0$, $\Delta p \rightarrow$ infinity, the x-quantum states become essentially continuous and the limit has been dubbed the "continuous variable" (CV) limit or description of quantum mechanics—a misnomer as "continuous" in the mathematical sense is contradictory to quantum physics that embodies the discrete nature of observables. The quantum mechanical states in the CV description are dubbed "qumodes" to distinguish from the true discrete nature of quantum mechanical states that carry information as "qubits".

There exist thus two approaches to QIP, one utilizing discrete variables and another that utilizes continuous variables. For either descriptive approach, Boson Sampling (BS) [4,5,6] is an approximate way and alternative to computing that allows simulating Bosonic quantum systems. The "Boson Sampling" approach to *simulate* Bosonic quantum systems discussed here is for N-photon states (Fock states) created from single photons emitted *on-demand* in a horizontal propagating mode that is interconnected to many (M-1) others in an on-chip horizontal network configuration compatible with realizing *at-scale* on-chip quantum optical circuits designed to create networks that provide the desired high -level function emerging from the functions of the individual building blocks [Fig.1]. Currently, the more commonly employed approach to photon-based Boson sampling is the use of Gaussian photon states [7] generated using readily available photon sources (typically an external laser) though not necessarily the best suited for on-chip QIP.

Thus, Boson sampling falls in one of two prominent classes: (1) Fock state-based Boson sampling, and (2) Gaussian boson sampling. The Fock state-based boson sampling is where actual quantum states of light are used as resource, and it corresponds to the original idea of computational complexity in linear optics addressed by Aaronson and Arkhipov in 2011 [5]. The computational complexity arises from the large number of permutations needed to symmetrize a multi-photon

Fock states [5]. It was soon realized that the probabilistic nature of the readily available parametric down conversion as a source for single photons does not allow scalable demonstration of Boson sampling. Thus, the Gaussian Boson Sampling (GBS) was proposed as a way-out the lack of scalable on demand single photon sources in 2017 by Hamilton et. al [7]. They showed that the output probability distribution of the photons can be nontrivial to compute classically in polynomial time with squeezed modes as inputs. However, since many squeezed states can be created readily using array of nonlinear optical waveguides, GBS was adopted as the go-to approach to demonstrate computational complexity in linear optics with large scale (~100s of modes) circuits. The few experimental works that employ epitaxial quantum dot as on-demand single photon source-based boson sampling have shown enormous advantage in terms of sampling rate [8,9]. However, large scale Fock state-based boson sampling has been out of reach because of the obvious limitation of the need for single photon source array.

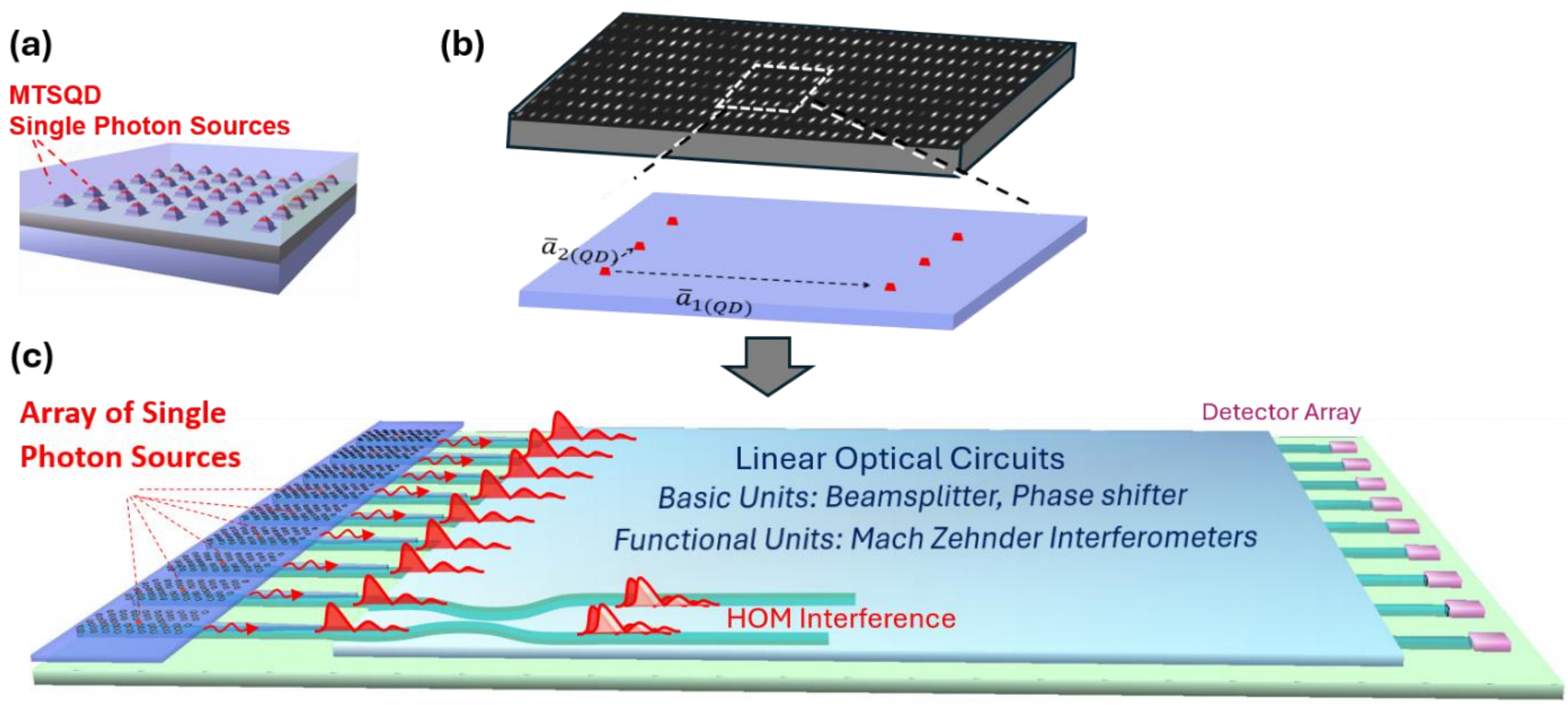

**Figure 1.** MTSQD array [red dots in panel (a) schematic & panel (b) white dots a large area photoluminescence image] as the enabler [panel (c) schematic] for standard Fock state-based Boson sampling in large arrays that offer better computational complexity than the Gaussian Boson sampling.

However, realization of linear optical quantum optical circuit has been made possible by the development of regular arrays of spectrally uniform quantum dots array, dubbed mesa top single quantum dot (MTSQD) array that are naturally amenable to chip-scale integration and provide single photons on-demand with spectral purity and visibility required for creating multi-photon entanglement [1,2,10]. Al illustrative example of this is shown in Fig. 2, wherein many MTSQD SPSs integrated with on-chip optical circuits can provide N-photon deterministic resource state which then can be manipulated in the optical circuit to enable entanglement. The single photon qubits from MTSQDs are shown to be high purity (>99.5%) with high visibility (>95%) at 4K. The two-qubit (photon) CNOT gate and Bell sates generated from such MTSQD single qubits exhibit, respectively, ~90% and ~82% fidelity [2]. While optimization of the characteristics of MTSQDs as quantum emitters are on-going, such on-demand quantum emitters in designed spatially known locations to an accuracy of better than ~2nm provide, for the first time, a meaningful starting platform for fabricating multi-photon networks / circuits and studying their decoherence dynamics. The natural question however is, how the less-than-ideal physical properties of the quantum emitter, in this case MTSQD, producing the multi-photon resource state affect the measurement outcome. Since most of the development on Boson sampling has been either assuming perfect Fock states, or with Gaussian squeezed states, the design principles of Boson sampling with real on-chip single photon sources remain essentially unexplored.

In this work we report a framework to analyze interference and *propagation of photon wave packets as representative of the single photons* emitted from a realistic photon source. Specifically, we address incorporating physical characteristics such as radiative decay lifetime ($T_1$), pure dephasing ($T_2^*$), and self-interference effect resulting in modulation of the photon wave packet intensity, into the output probability distribution that arise because of the multi-photon interference. Note that these effects have only accounted for two photon interference in earlier works [1,11,12]. We report a scheme to efficiently account for interference of multiple photon wave packets of arbitrary shapes, with full accounting of the symmetrization of the photons in various modes. The analysis reported here takes MTSQD array as an example of technologically viable and established way of realization, but we note that the concept of understanding effect of photon wave packet's shape is equally applicable to any single photon emitter.

## II. Theoretical Framework: Multi-photon Interference for Photon Wave packets

### II.A. Wave packet description of multi-photon resource state

To faithfully represent the resource state comprising of single photons emitted from an array of single photon emitters, three timescales need to be accounted for: (1) The energy ($\omega$), (2) The time-span of the wave packet given by the radiative decay lifetime of the emitter ($T_1$), (3) and the pure dephasing rate ($\gamma_\phi$) representing the rate of introduction of phase uncertainty due to interaction of the single photon emitter with environment during the emission process. Important to note, the finite lifetime of the photon wave packet means that the energy is no longer well-defined quantum number, and thus the wave packet can be expressed as a superposition of the basis of perfectly periodic waves. Thus, any single photon state emitted by a single photon emitter has finite width in both time and in energy.

We can readily represent such a single photon state using a Fourier series as:

$$|\Psi_{WP}\rangle = \int d\omega \tilde{f}(\omega)\, a_\sigma^\dagger(\omega)|0\rangle = \int_0^\infty dt f(t)\, a_\sigma^\dagger(t)|0\rangle \tag{1}$$

Here the functions $\tilde{f}(\omega)$ and $f(t)$ represents the shape of the wave packet in energy and time domain respectively. These functions are Fourier transforms of each other.

The shape of the wave packet depends on the nature of light matter coupling between the single photon source and the output photon mode. In most photonic devices, including cavity-coupled SPSs, the weak-coupling limit can be assumed which simplifies the contribution of the vacuum fluctuation as a single exponential decay term, characterized by the radiative decay lifetime $T_1$, and an oscillating part characterized by the transition energy of the emitter, $\omega_0$. In this regime, the wave packet shape function $f(t)$ can be readily expressed in terms of physical parameters as $f(t) = e^{-i\omega_0 t} e^{-t\,/\,2T_1} e^{-i\phi(t)}$.

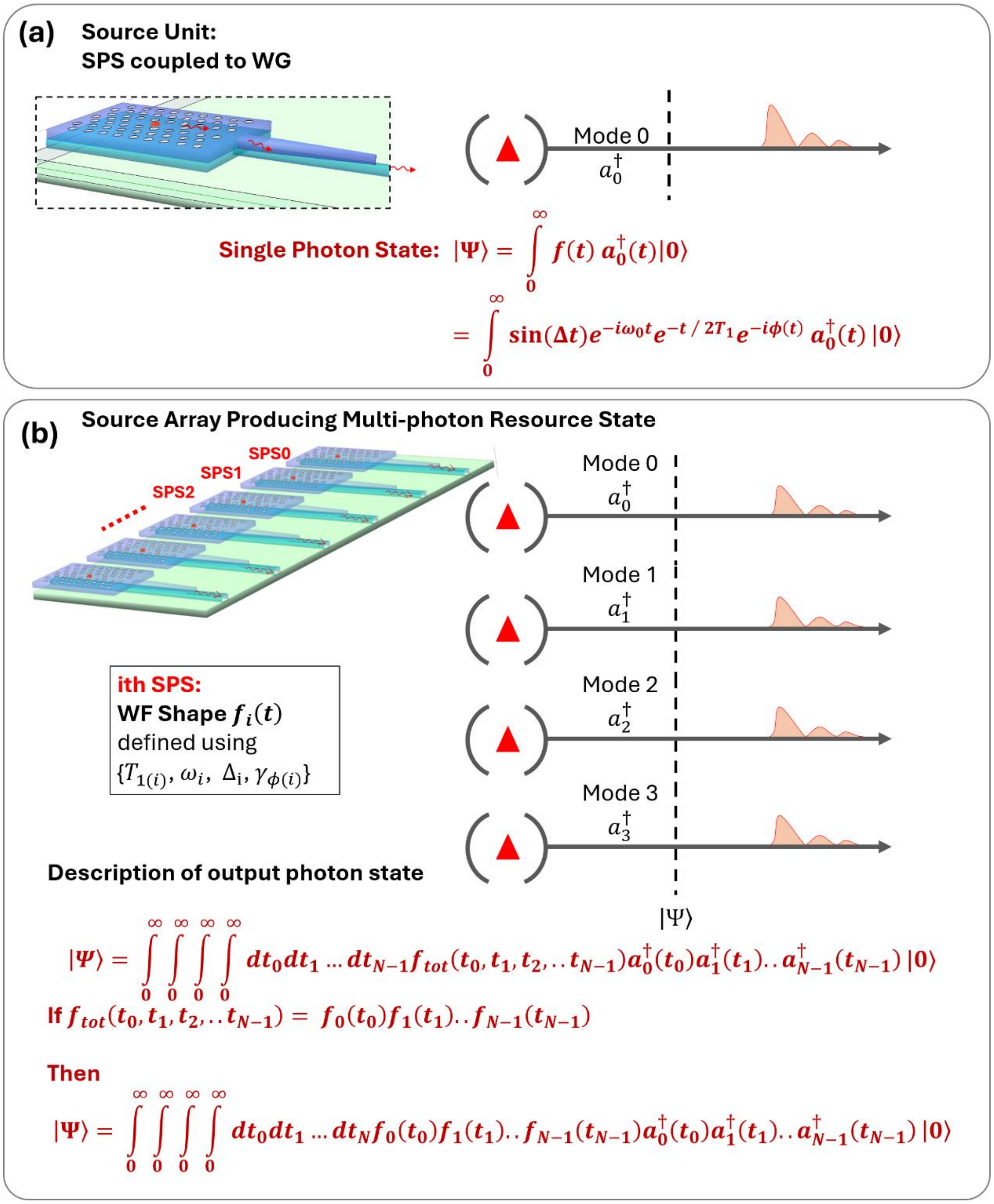

**Figure 2.** Description of single and multi-photon resource state (a) Single Photon wave packet produced from a single MTSQD SPS coupled to photonic cavity and waveguide. The wave packet can be characterized by $T_1$ (radiative decay lifetime), $\Delta$ (fine structure splitting, and $\omega$ (energy). (b) Description of a N-photon resource state that can be produced by an ordered array of SPSs (such as the MTSQDs) in the wave packet representation.

Here $\phi(t)$ is a random phase noise resulting from the accumulation of the random energy shift F(t) and is given by,

$$\phi(t) = \int_0^t F(t')dt' \tag{2}$$

In the simplest approximation where the environment has no memory, $F(t)$ is a delta-function correlated Gaussian random variable, characterized by,

$$\ll F(t)F(t') \gg = 2\gamma_\phi \delta(t - t') \tag{3}$$

Here $\ll\cdot\gg$ represents statistical average of the energy fluctuation time autocorrelation. The pure dephasing rate is $\gamma_\phi$. For the moment we assume it is a constant. Of course, as seen in recent works [2] the pure dephasing rate can be a function of time if spectral diffusion processes arising from residual charge noise are involved. Like other epitaxial quantum dots, the MTSQD single photon source can emit the single photons from a superposition of finely separated levels resulting in a self-interference beating pattern in the wave packet shape [1]. Also, using strongly coupled cavity mode, Rabi oscillation can significantly change the nature of $f(t)$ from a simple exponential function as seen in the Wigner-Weisskopf approximation. Thus, for translatability of the theoretical framework to real devices, we keep the generic form of $f(t)$ and we account for it numerically in a rotating basis of $\omega_0$, thus discarding rapidly oscillating terms.

Importantly the wave packet description in equation (1) can be extended to a N-photon resource state generated by a N-SPSs in an array, as schematically depicted in Fig.1, MTSQD regular array single photon sources provide the natural platform to realize such massively parallel N-photon deterministic resource state generation. The Fock basis we represent as $\bar{n} = (n_1, n_2, \dots, n_N)$, where $n_i$ represents the number of photons in the $i$ th waveguide. The overall photon state created by the N single photon sources are thus given by $\bar{n} = (1,1,1,1, \dots)$, i.e. one photon in each waveguide mode. Thus, the overall N-photon wave packet can be represented as

$$|\Psi\rangle = \int_0^\infty \int_0^\infty \int_0^\infty \dots \int_0^\infty dt_0 dt_1 \dots dt_{N-1} f_{tot}(t_0, t_1, t_2, \dots t_{N-1}) a_0^\dagger(t_0) a_1^\dagger(t_1) \dots a_{N-1}^\dagger(t_{N-1}) \,|0\rangle \tag{4}$$

in which $f_{tot}(t_0, t_1, t_2, \dots t_{N-1})$ represents the wave packet of the N-photons. In certain cases, it may be well approximated as

$$f_{tot}(t_0, t_1, t_2, .. \ t_{N-1}) \equiv \ f_0(t_0) f_1(t_1) .. f_{N-1}(t_{N-1}) \tag{5}$$

so that

$$|\Psi\rangle = \int_0^\infty \int_0^\infty \int_0^\infty \dots \int_0^\infty dt_0 dt_1 \dots dt_N f_0(t_0) f_1(t_1) .. f_{N-1}(t_{N-1}) a_0^\dagger(t_0) a_1^\dagger(t_1) .. a_{N-1}^\dagger(t_{N-1}) \, |0\rangle \tag{6}$$

is a reasonable representation of the N-photon wave packet

**II.B. Symmetrization of the N-photon State**

As per the rules of Bose statistics the N photon state needs to be always symmetrized, i.e. invariant under swap of any two photons. Since the Bosonic creation operators are assumed to obey the commutation relation $[a_i, a_j^\dagger] = \delta_{ij}$, the state shown in equation (6) is automatically symmetrized. It can be also explicitly symmetrized by expressing it as a sum over product states, as,

$$|\Psi\rangle$$

$$= \int_0^\infty \int_0^\infty \int_0^\infty \dots \int_0^\infty dt_0 dt_1 \dots dt_{N-1} f_{tot}(t_0, t_1, t_2, .. \ t_{N-1}) \, |1_0(t_0), 1_1(t_1), \dots, 1_{N-1}(t_{N-1})\rangle$$

$$= \hat{S} \left[ \int_0^\infty \int_0^\infty \int_0^\infty \dots \int_0^\infty dt_0 dt_1 \dots dt_{N-1} f_{tot}(t_0, t_1, t_2, .. \ t_{N-1}) \, |1_0(t_0)\rangle |1_1(t_1)\rangle .. |1_{N-1}(t_{N-1})\rangle \right] \tag{7}$$

In this notation, we represent n photon occupying mode $i$ at time $t$ as $|n_i(t\,)\rangle$. Here the operator $\hat{S}$ sums over all possible permutations given by $\hat{S} = \frac{1}{\sqrt{N!}} \sum_{P \in S_N} \hat{P}$ with $S_N$ being the permutation group for N photons with assuming no two photons in the same mode. Here we also use the fact that each photon occupies a distinct non-overlapping mode at the start so that we do not need to consider the symmetrizing of multiple photons occupying the same mode. Up to now all the photons are in different modes, hence their creation operator commutes. That is

$$[a_i, a_j^\dagger] = \delta_{ij} \tag{8}$$

The commutator vanishes for photons in different modes. However, as soon as we have multiple photons in the same waveguide mode, we will need symmetrization between photons that are in the same mode. This is done in Section II.F. Before incorporating this explicit symmetrization, we first do some notational simplification.

We write,

$$f_{tot}(0,1,2,3, \dots N-1) \equiv f_{tot}(t_0, t_1, t_2, .. \ t_{N-1}) \equiv \ f_0(t_0) f_1(t_1) .. f_{N-1}(t_{N-1}) \tag{9}$$

From here onwards we can treat $f$ as a N-dimensional matrix where each dimension is representative of the time coordinate of the photons emitted from the single photon source array. We also simplify how the state is represented as $a_i^\dagger(t_j)|0\rangle \equiv |j_i\rangle$. Then the symmetrized Fock state can be written as

$$|\Phi_{\{1,1,1,1..\}}\rangle = a_0^\dagger(t_0)a_1^\dagger(t_1)..a_{N-1}^\dagger(t_{N-1})|0\rangle \equiv |0_0 1_1 \dots (N-1)_{N-1}\rangle \tag{10}$$

The unsymmetrized product state of N-photons each occupying a single mode is represented as

$$|1_0(t_0)\rangle|1_1(t_1)\rangle..|1_{N-1}(t_{N-1})\rangle = |0_0\rangle|1_1\rangle..|(N-1)_{N-1}\rangle \tag{11}$$

With this notation, we write the initial N-photon resource state in equation (7) as

$$|\Psi\rangle = \int d(0,1,2\dots)\, f_{tot}(0,1,2\dots)\, |0_0 1_1 2_2 \dots\rangle$$

$$= \hat{S}\left[\int d(0,1,2\dots)\, f_{tot}(0,1,2\dots)\, |0_0\rangle|1_1\rangle|2_2\rangle \dots\right] \tag{12}$$

**II.C. Photon Interference:**

Now evaluate interference of a multi-photon wave packet. Figure 3 represents a simple circuit to highlight the procedure. An example is shown in Figure 3.

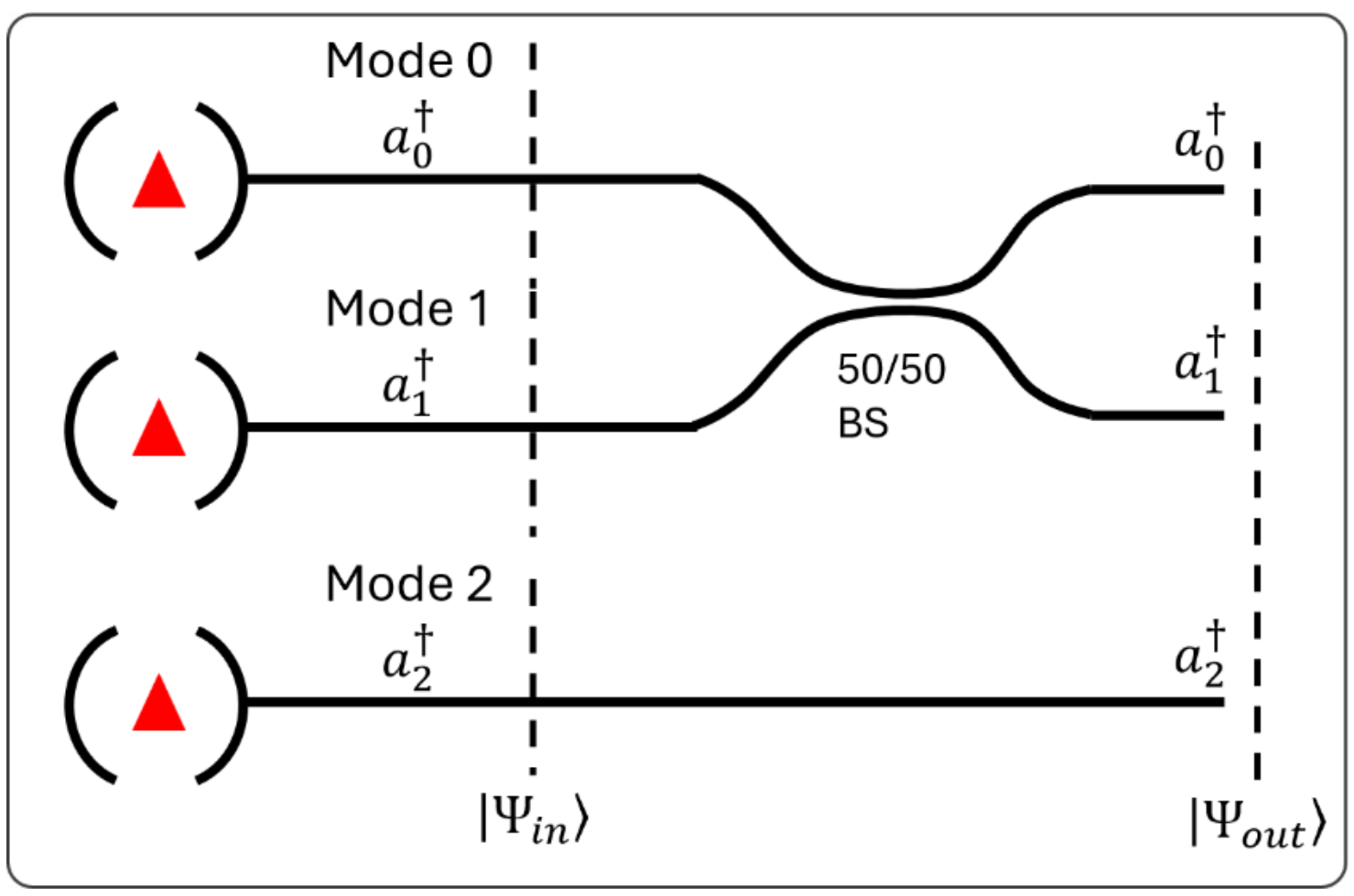

**Figure 3.** Example of multiphoton wave packet transformation by a simple linear optical circuit in a 3 waveguide, 3 photon configurations.

As shown in Figure 3, we illustrate with a 3-photons emitted into a 3-waveguide optical circuit, where the three single photon sources create a starting resource state in the Fock basis of $\bar{n} =$ (1,1,1) (i.e. 1 photon in each waveguide). As per the convention set up in equation (12), this 3-photon starting state can be represented as

$$|\Psi_{in}\rangle = \int d0d1d2\, f_{tot}(0,1,2)\ |0_0 1_1 2_2\rangle$$
$$= \hat{S}\left[\iint d0d1d2\, f_{tot}(0,1,2)\ |0_0\rangle|1_1\rangle|2_2\rangle\right] \tag{13}$$

The beamsplitter added in the circuit shown in figure 5 mixes the modes 0 and 1. The beamsplitter transformation in the usual full description is represented as

$$a_0^\dagger \rightarrow \frac{1}{\sqrt{2}}\left[ia_0^\dagger + a_1^\dagger\right]\ and\ \ a_1^\dagger \rightarrow \frac{1}{\sqrt{2}}\left[ia_1^\dagger + a_0^\dagger\right] \tag{14}$$

In our simplified notation equation (17) is written as

$$|0_0\rangle \rightarrow \frac{1}{\sqrt{2}}[i|0_0\rangle + |0_1\rangle], |1_1\rangle \rightarrow \frac{1}{\sqrt{2}}[i|1_1\rangle + |1_0\rangle] \tag{15}$$

Then the final state after transformation by the beamsplitter is represented as

$$|\Psi_{out}\rangle = \frac{1}{2}\hat{S}\left[\iint d0d1d2\, f_{tot}(0,1,2)\ [-|0_0\rangle|1_1\rangle|2_2\rangle + |1_0\rangle|0_1\rangle|2_2\rangle + i|0_1\rangle|1_1\rangle|2_2\rangle + i|0_0\rangle|1_0\rangle|2_2\rangle]\right] \tag{16}$$

**II.D. Explicit Accounting for Photon Wavefunction Symmetrization:**

In this section, we take the most important step to allow computation for general optical circuits. We take the burden of symmetrization from the states to the wave packet shape function f. In the following I present a simple illustration:

Let us say we have a state $\psi = |0_0 1_1 2_2\rangle$. This state means that the photon with time index 0 ($t_0$) resides in mode 0, photon with time index 1 ($t_1$) resides in mode 1 and so on. Because all the bosonic operators of different modes commute, i.e. $\left[a_0, a_1^\dagger\right] = \left[a_1, a_2^\dagger\right] = \left[a_2, a_0^\dagger\right] = 0$, the symmetrization, denoted by $\hat{S}$ has no impact on the observable. The photons can be treated as independent and far apart without any consequence.

On the other hand, if a mode has two photons, they need to be explicitly symmetrized.

Thus, $f_{tot}(0,1,2)|0_0 1_0 2_2\rangle = \frac{1}{\sqrt{2}}\ [f_{tot}(0,1,2)|0_0\rangle|1_0\rangle|2_2\rangle + f_{tot}(0,1,2)\ |1_0\rangle|0_0\rangle|2_2\rangle]$

Now, because the time coordinates are integrated in equation (16), we can change the labels of the time coordinates resulting in a change in the ordering of the time axes in the function f. Practically this means that the output state of the circuit in Figure 3, expressed in equation (16) can be expressed as,

$$|\Psi_{ou}\rangle = \frac{1}{2}\hat{S}\int d0d1d2\, f_{tot}(0,1,2)\ \ [-|0_0\rangle|1_1\rangle|2_2\rangle + |1_0\rangle|0_1\rangle|2_2\rangle + i|0_1\rangle|1_1\rangle|2_2\rangle + i|0_0\rangle|1_0\rangle|2_2\rangle]$$

$$= \frac{1}{2}\hat{S}\int d0d1d2[-f_{tot}(0,1,2) + \ f_{tot}(1,0,2)]\ |0_0\rangle|1_1\rangle|2_2\rangle$$

$$+ \frac{i}{2\sqrt{2}}\hat{S}\int d0d1d2[f_{tot}(0,1,2) + \ f_{tot}(1,0,2)]\ |0_0\rangle|1_0\rangle|2_2\rangle$$

$$+ \frac{i}{2\sqrt{2}}\hat{S}\int d0d1d2[f_{tot}(0,1,2) + \ f_{tot}(1,0,2)]\ |0_1\rangle|1_1\rangle|2_2\rangle \qquad (17)$$

Equation (17) is identical to equation (16) but it is an enormous simplification in calculating the final state and emulating photon counting measurements. Here the exchange between the photons is removed, and instead, the transposition of the matrix $f$ between various time-dimensions serves as to capture the effect of photon exchange.

**II.E. Simplified description of input and output of the circuit shown in Fig. 5 and definition of a Transfer Matrix.**

From this point onwards we think of the function $f_{tot}(0,1,2)$ as a 3-dimensional matrix, each dimension representing a time axis. Rearranging the time labels is equivalent to transposing the matrix along various combinations of axes. For example, a swap between photon # 0 and photon # 1 is represented by a transposition along the first two axes, i.e. $f_{tot}(0,1,2) \rightarrow \ f_{tot}(0,1,2)$. The burden of symmetrization is taken away from the state, and it is imposed in these wave packet shape matrices denoted by $f$.

The initial state, in equation (16) represents a Fock state $\bar{n}_{in} = (1,1,1)$, and the output states shown in equation (19) and (20) contain Fock states (1,1,1), (0,2,1), and (2,0,1). For each pair of Fock state, a transformation can be defined for each set of permutations. Thus, the application of the circuit can be shorthand represented as

$$f_{tot}(0,1,2)\Phi_{(1,1,1)}$$
$$\rightarrow \left[\frac{1}{2}\ f_{tot}(1,0,2) - \frac{1}{2}f_{tot}(1,0,2)\right]\ \Phi_{(1,1,1)}$$
$$+\ \frac{i}{2\sqrt{2}}[\ f_{tot}(1,0,2) + f_{tot}(1,0,2)]\Phi_{(0,2,1)}$$
$$+\ \frac{i}{2\sqrt{2}}[\ f_{tot}(1,0,2) + f_{tot}(1,0,2)]\Phi_{(2,0,1)} \qquad (18)$$

where $\Phi_{(1,1,1)} = \ |0_0\rangle|1_1\rangle|2_2\rangle$. In general, $\Phi_{(m,n,p)} = \ |0_0\rangle|1_0\rangle|2_0\rangle..|m-1_0\rangle\ |m_1\rangle|m+1\ _1\rangle..|m+n-1\ _1\rangle|m+n\ _2\rangle|m+n\ +1_2\rangle \ldots\ |m+n\ +p-1_2\rangle$, i.e. the unsymmetrized photon product state of m photons in mode 0, n photons in mode 1, and p photons in mode 2. These are a basis but

not physical states, and their use makes sense when the symmetrization is explicitly accounted in the weight functions ($f$) as incorporated in (18) above.

We can generalize equations (18) by defining a transition matrix that transforms a wave packet $f_{tot}(0,1,2,\dots,N)$ corresponding to a mode $\bar{n} = (n_1, n_2, \dots., n_N)$for a $N$ mode optical circuit with a total photon number of N= sum($\bar{n}$) $= \sum n_i$. Let us denote the transfer matrix by $T$. Then the equation (18) can be generalized for arbitrary number of photon wave packets as:

$$[f]\,\Phi_{\bar{n}} \rightarrow \sum_{\bar{m}\,,\sigma\in P_M} T_{\bar{m},\bar{n},\sigma}\;\sigma[f]\,\Phi_{\bar{m}} \tag{19}$$

Where, $[f]$ is representative of the N dimensional waveform shape matrix where each axis is representative of the time coordinate of each of the M photons. Here $\bar{n}$ and $\bar{m}$ denote the Fock state basis which uniquely tells how many photons are in which modes. For total N photons, distributed in total N waveguides, we have $\binom{M+N-1}{M-1}$ ways of choosing the Fock states. Thus $n_{dim} = \binom{M+N-1}{M-1}$ represents the size of the first two axes of the T matrix. Finally, $P_N$ is the set of all permutations of the N time axes, and $\sigma$ represents a specific permutation in this set which acts on the f matrix and shuffles the time axes. Clearly the total number of permutations in this case is $N!$. The total number of elements in the transformation tensor T is thus $n_{dim} \times n_{dim} \times N! = \binom{M+N-1}{M-1}^2 N!$. This is an exponentially increasing function with number of photons N, and it also increases exponentially with increasing modes (M) for large N. Figure 4 shows the total dimension $n_{dim} \times n_{dim} \times N!$ plotted with respect to number of photons and number of modes, demonstrating the steep increase in computational complexity – the characteristics of Boson sampling.

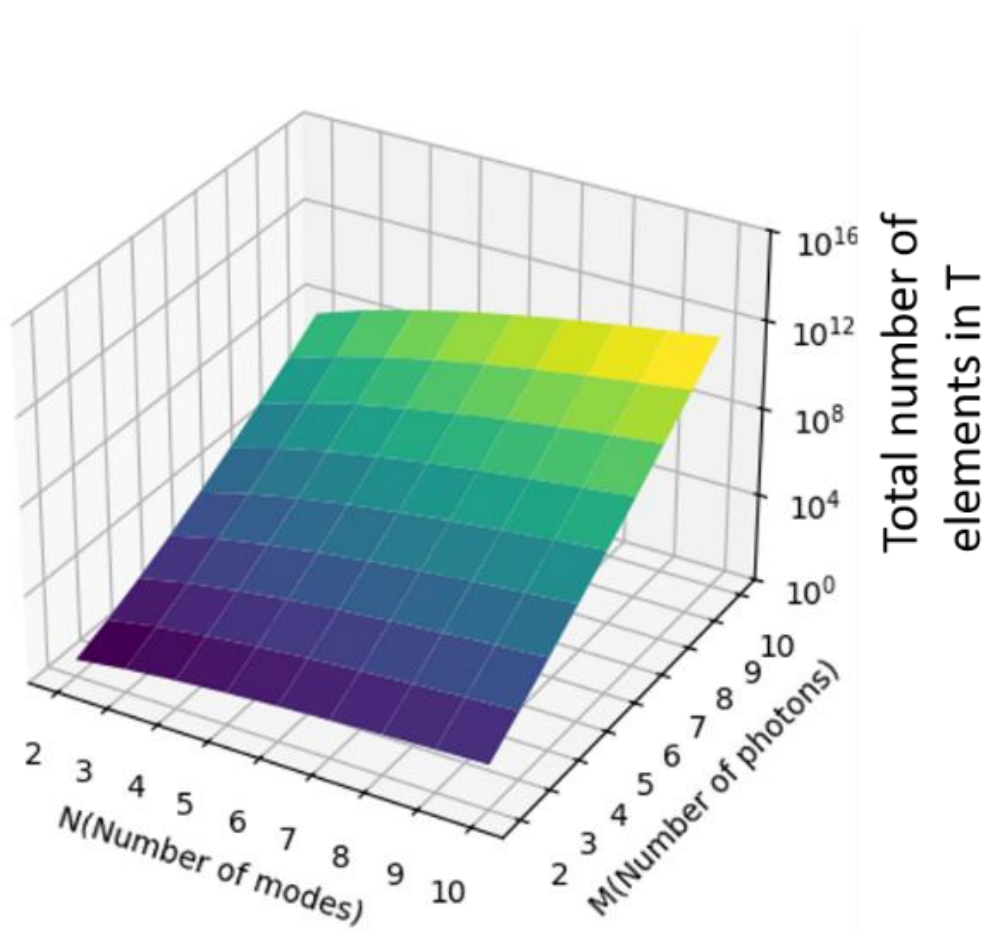

**Figure 4.** Total number of elements of the T matrix, $\binom{M+N-1}{M-1}^2 N!$, plotted against N (number of photons) and M (number of modes). A brute force approach to evaluate T for ~10 photons in ~10 modes is clearly impossible.

## III. Results

Based on the above framework, we now address how does the pure dephasing that each photon carries in its phase noise affects the photon correlation measurements for M-photon N-mode circuit in general.

### A. Derivation of Photon Coincidence Count Histogram from the Output State

As an initial example situation, we assume that each photon is identical except only for the pure dephasing part. Thus,

$$f_i(t) = e^{-i\omega_0 t} e^{-t\,/\,2T_1} e^{-i\phi_i(t)} \;\forall\, i \tag{20}$$

Of course, the variability of the physical parameters between the MTSQDs in the array is important for large scale implementation in real QIP systems of interest, and its effect on the correlation measurement outcomes will be addressed in a future work. We also assume that the phase noises of different sources are uncorrelated. That is

$$\ll e^{-i\phi_i(t)} e^{-i\phi_j(t)} \gg = \ll e^{-i\phi_i(t)} \gg \ll e^{-i\phi_j(t)} \gg \quad \forall\, i \neq j \tag{21}$$

Physically this approximation corresponds to the noise that the multiple MTSQDs are subjected to being independent. In reality, while phonon bath induced noise can be assumed independent in general due to its highly incoherent nature, for electrical and charge noise one may expect a correlation between neighboring SPSs in which case equation (21) will not be strictly valid. In this work, however, we limit ourselves to the simplest approximation of uncorrelated phase noise. We recall that for purely gaussian and delta-correlated dephasing process

$$\ll e^{-i\phi(t_1) + i\phi(t_2)} \gg = e^{-\gamma_\phi |t_1 - t_2|} \tag{22}$$

Equations (20), (21), and (22) along with the axis reordering recorded via $T_{\bar{m},\bar{n},\sigma}$ matrix formulation can be now used to predict exactly the shape of the coincidence count histogram measurement.

Let us take a concrete example of the circuit shown on Fig. 3. We recall from equation (16) and (17) that the output state is given by

$$\begin{aligned}
|\Psi_2\rangle &= \frac{1}{2}\int d0d1d2\; f_{tot}(0,1,2)\;\; [-|0_0 1_1 2_2\rangle + |1_0 0_1 2_2\rangle + i|0_1 1_1 2_2\rangle + i|0_0 1_0 2_2\rangle] \\
&= \frac{1}{2}\int d0d1d2[-f_{tot}(0,1,2) + f_{tot}(1,0,2)]\,|0_0\rangle|1_1\rangle|2_2\rangle \\
&+ \frac{i}{2\sqrt{2}}\int d0d1d2[f_{tot}(0,1,2) + f_{tot}(1,0,2)]\,|0_0\rangle|1_0\rangle|2_2\rangle \\
&+ \frac{i}{2\sqrt{2}}\int d0d1d2[f_{tot}(0,1,2) + f_{tot}(1,0,2)]\,|0_1\rangle|1_1\rangle|2_2\rangle
\end{aligned} \tag{23}$$

which can be expressed in shorthand notation, repeated below

$$|\Psi_2\rangle = \left[\frac{1}{2} f_{tot}(1,0,2) - \frac{1}{2} f_{tot}(1,0,2)\right] \Phi_{(1,1,1)} + \frac{i}{2\sqrt{2}}[\, f_{tot}(1,0,2) + f_{tot}(1,0,2)]\Phi_{(0,2,1)} + \frac{i}{2\sqrt{2}}[\, f_{tot}(1,0,2) + f_{tot}(1,0,2)]\Phi_{(2,0,1)} \qquad (24)$$

Now, let us say that we measure 1 photon in detector 0 at time $t_{M1}$ and 1 photon in detector 1 at time $t_{M2}$. We assume that the circuits are equipped with photon number resolving detectors for this example. We denote the coincidence count histogram by $g_{01}^{(2)}(t_{M1}, t_{M2})$. From the state shown in equation (24) above, we can say that

$$g_{01}^{(2)}(t_{M1}, t_{M2}) = \int dt \left|\frac{1}{2}[- f_{tot}(t_{M1}, t_{M2}, t) + f_{tot}(t_{M2}, t_{M1}, t)]\right|^2 \qquad (25)$$

The time dimension of the third photon at detector 2 in this case is simply integrated out.

Then we have

$$g_{01}^{(2)}(t_{M1}, t_{M2}) = \frac{1}{4}|f(t_{M1})|^2|f(t_{M2})|^2 \ll 2 - e^{-i\phi_0(t_{M1})+ i\phi_0(t_{M2})}e^{-i\phi_1(t_{M1})+ i\phi_1(t_{M2})} - e^{i\phi_0(t_{M1})- i\phi_0(t_{M2})}e^{i\phi_1(t_{M1})- i\phi_1(t_{M2})} \gg \qquad (26)$$

By invoking statistical independence of phase noise of first photon ($\phi_0$) with the phase noise of the second photon ($\phi_1$), we have

$$g_{01}^{(2)}(t_{M1}, t_{M2}) = \frac{1}{4}|f(t_{M1})|^2|f(t_{M2})|^2\left[2 - 2 \ll e^{-i\phi_0(t_{M1})+ i\phi_0(t_{M2})} \gg\ll e^{-i\phi_1(t_{M1})+ i\phi_1(t_{M2})} \gg\right] \qquad (27)$$

By invoking equation (22) above, we have

$$g_{01}^{(2)}(t_{M1}, t_{M2}) = \frac{1}{4}|f(t_{M1})|^2|f(t_{M2})|^2\left[2 - 2e^{-2\gamma_\phi|t_{M1}-t_{M2}|}\right] = \frac{1}{2}|f(t_{M1})|^2|f(t_{M2})|^2\left[1 - e^{-|t_{M1}-t_{M2}|/T_2^*}\right] \qquad (28)$$

This is an exact expression of the 2 -photon coincidence count histogram as reported in Ref. [1].

### B. Four-photons in four-modes

The above approach is general. We can use the same framework to arrive at the coincidence count histogram for any arbitrary number of photons detected. To demonstrate, in Fig. 5 (a) and (b), we show a simple circuit beyond the single HOM interference [11,12], with 4 photon resource state and 4 photonic modes in parallel. As evident from the structure, the first stage of beamsplitters creates a pair of 2 photon NOON state at the middle of the circuit, and then the two 2-photon NOON states interfere in the second beamsplitter producing a 4-photon NOON state [13].

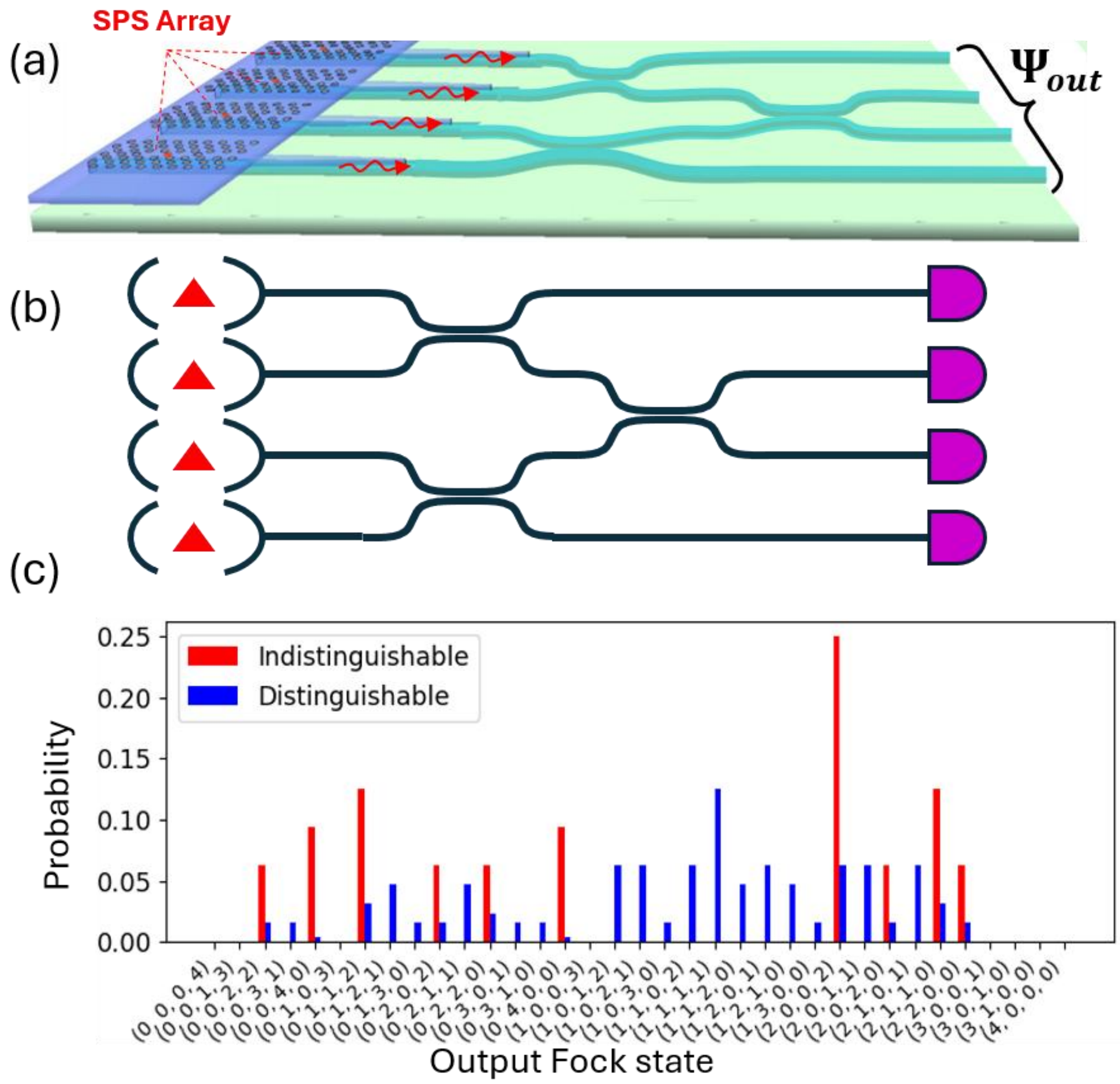

**Figure 5**. **(a)** Device level schematic and **(b)** circuit-schematic of the 4-mode 4-photon QIP circuit and **(c)** the probability distribution of various Fock state configurations at the output. The red bars assume perfectly indistinguishable photons and the blue bars indicate distinguishable photons/ classical light.

**Output probability distribution—distinguishable vs indistinguishable, --**

In Fig. 5(c) we show the probability of measuring various Fock states at the output. Here the red bars indicate the case where each photon is assumed to be perfectly identical, and the blue bars represent the case where the photons are fully incoherent with each other, i.e. the classical limit. The difference between these two extremes signifies the bunching behavior of indistinguishable photons. As an example, the 4-photon NOON state $\frac{1}{\sqrt{2}}(|0,0,4,0\rangle + |0,4,0,0\rangle)$ is produced with probability ~0.187 for indistinguishable photons, and its probability for distinguishable photons is greatly diminished ($\sim \frac{1}{256}$). On the other hand, the output state $|1,1,1,1\rangle$ is produced with probability $\frac{1}{8}$ for distinguishable photons, but its probability is fully suppressed for indistinguishable photons. Note that, to obtain these extreme cases of perfectly indistinguishable vs distinguishable photons, it is not necessary to account for the photon wave packet shapes. They can be obtained by assuming ideal Fock states at the input, for example $|\Psi_{in}\rangle = a_0^\dagger a_1^\dagger a_2^\dagger a_3^\dagger|0\rangle$, and imposing appropriate commutation relation for indistinguishable and distinguishable photons [4,5].

**Effect of dephasing on probability, --**

In reality, the single photons are not perfectly indistinguishable Fock states. They are characterized by several parameters that have direct physical relevance to the electronic structure of the SPS and the noise sources that impact the SPS. For example, in Fig. 6(a) we show the single photon wave packet shape ($|f(t)|^2$) measured for the MTSQD generated single photon state which is characterized by a radiative decay lifetime $T_1 = 0.35ns$, a beating pattern arising from a fine structure splitting of $\Delta_{FSS} = 6.4\mu eV$. Also, the dephasing rates are well characterized for this class of MTSQD SPSs [1]. Figure 6(b) shows the effect of the temperature dependent dephasing, on the two-photon HOM interference visibility ($V_{TPI}$), indicating >95% visibility at LHe temperature. Note that $V_{TPI}$ is determined from the relative timescale between the dephasing and decay, i.e. $V_{TPI} = \frac{1/2T_1}{1/2T_1 + 1/T_2^*}$ [1].

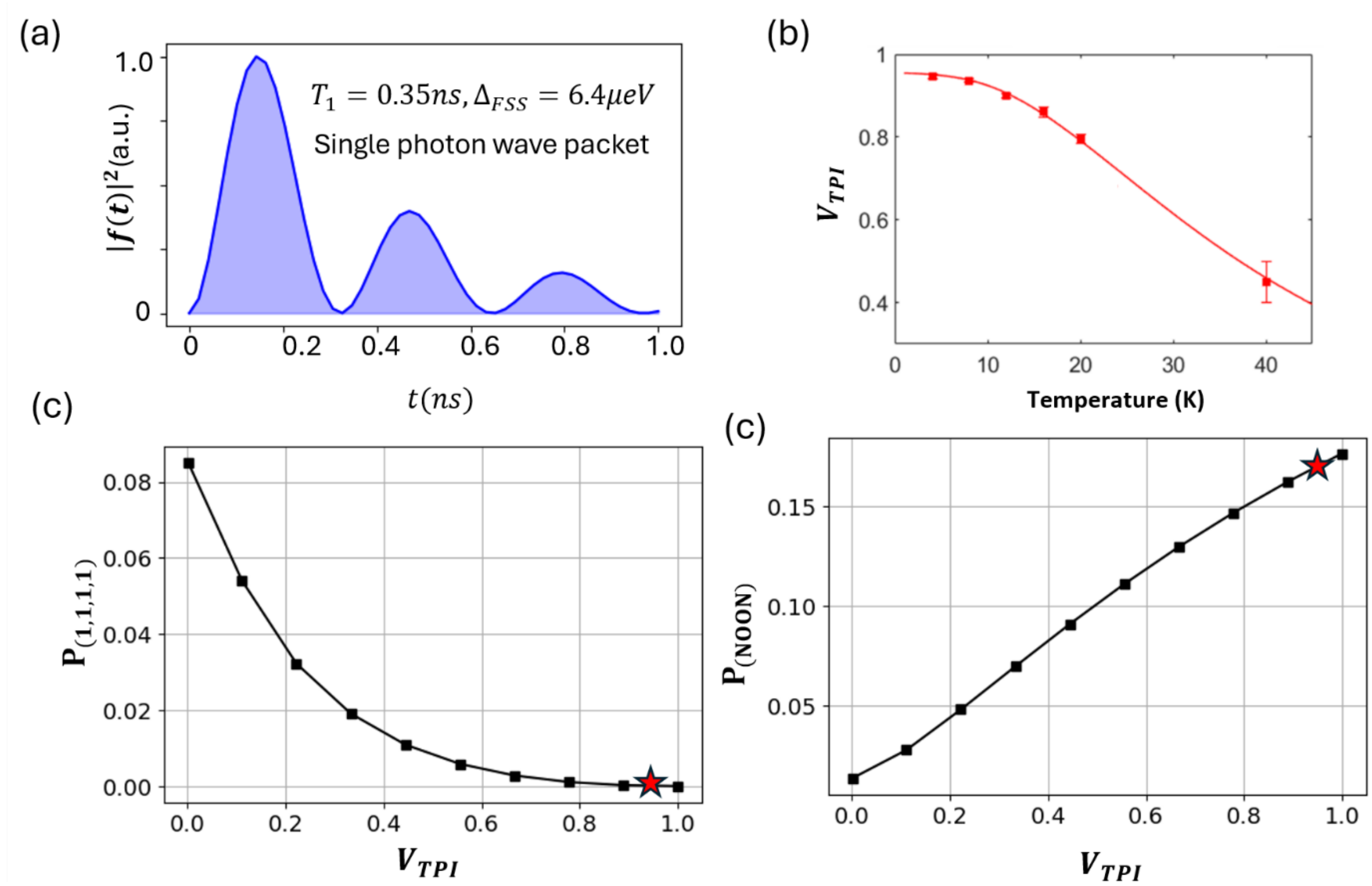

**Figure 6.** For the circuit shown in Fig. 5, we assume each photon has a wave packet shape shown in (a). Note, panel **(a)** represents the measured single photon wave packet from representative MTSQD SPS [1]. **(b)** The dephasing part, not captured in the amplitude shown in (a), is quantified using two photon

interference visibility ($V_{TPI}$). (b) $V_{TPI}$ as a function of temperature for the MTSQD SPSs [2] where at T<10K, >95% visibility is achieved without Purcell enhancement. With Purcell effect, $V$ can be boosted to near unity. (c) Shows the integrated $g^{(4)}$ count for the output state $|1,1,1,1\rangle$ **(d)** shows the equivalent for the output 4-photon NOON state $\left(\frac{1}{\sqrt{2}}(|0,0,4,0\rangle + |0,4,0,0\rangle)\right)$. The star indicates the measured values [2]

Here, we investigate how the physical nature of the single photon wave packet shown in Fig. 6(a) and (b) affect the output probabilities in the circuit shown in Fig. 5. The input here is the state

$$\Psi_{in} = \int d0d1d2d3\, f(0,1,2,3)\Phi_{(1,1,1,1)} \tag{29}$$

When propagated through the circuit via the linear transformations defined in equation (19) the output Fock state component corresponding to the $|1,1,1,1\rangle$ Fock state is:

$$\langle \Phi_{(1,1,1,1)} |\, \Psi_{out}\rangle = \frac{1}{8}\int d0d1d2d3 \begin{bmatrix} -f(0,1,2,3) + f(0,1,3,2) + f(0,2,1,3) - f(0,2,3,1) \\ + f(1,0,2,3) - f(1,0,3,2) - f(2,0,1,3) + f(2,0,3,1) \end{bmatrix} \tag{30}$$

Thus, the probability of the $|1,1,1,1\rangle$ state at the output can be written as

$$P_{(1,1,1,1)} = \left|\langle \Phi_{(1,1,1,1)} |\, \Psi_{out}\rangle\right|^2 \tag{31}$$

The probability can be evaluated as

$$\begin{aligned} P_{(1,1,1,1)} = \frac{1}{64}\int d0d1d2d3|f(0)|^2|f(1)|^2|f(2)|^2|f(3)|^2\big[&8 - 4e^{-2\gamma_\phi\tau_{01}} - 4e^{-2\gamma_\phi\tau_{20}} \\ &- 8e^{-2\gamma_\phi\tau_{12}} - 4e^{-2\gamma_\phi\tau_{13}} - 4e^{-2\gamma_\phi\tau_{23}} + 8e^{-\gamma_\phi(\tau_{31}+\tau_{12}+\tau_{23})} \\ &+ 8e^{-\gamma_\phi(\tau_{01}+\tau_{12}+\tau_{20})} - 8e^{-\gamma_\phi(\tau_{02}+\tau_{10}+\tau_{23}+\tau_{13})} + 4e^{-2\gamma_\phi(\tau_{01}+\tau_{23})} \\ &+ 4e^{-2\gamma_\phi(\tau_{20}+\tau_{13})}\big] \end{aligned} \tag{32}$$

where $\tau_{ij} = t_i - t_j$. Equation (32) provides a direct link between the dephasing rate $\gamma_\phi$ and the probability of generating the output Fock state in the circuit shown in Fig. 5. In Figure 6(c) we plot $P_{(1,1,1,1)}$ as a function of $V_{TPI}$. The red star indicates current performance metric established for the MT SQD SPS, but the methodology here applies to estimating the probability for single photons generated from any other single photon sources as well.

In a similar way, we also estimate the output state projected along the 4-photon NOON state, i.e. $\Phi_{NOON} = \frac{1}{\sqrt{2}}(|0,0,4,0\rangle + |0,4,0,0\rangle)$.

$$\langle \Phi_{NOON} |\, \Psi_{out}\rangle = 0.018\int d0d1d2d3 \left[\sum_{p\in S_4} f(p_0,p_1,p_2,p_3)\right] \tag{33}$$

where $p \in S_4$ is all permutations of the 4 time-axes. The probability, $P_{NOON} = |\langle \Phi_{NOON} | \Psi_{out} \rangle|^2$ is shown in Fig. 6(d) as a function of $V_{TPI}$, showing a transition between the distinguishable and indistinguishable limits.

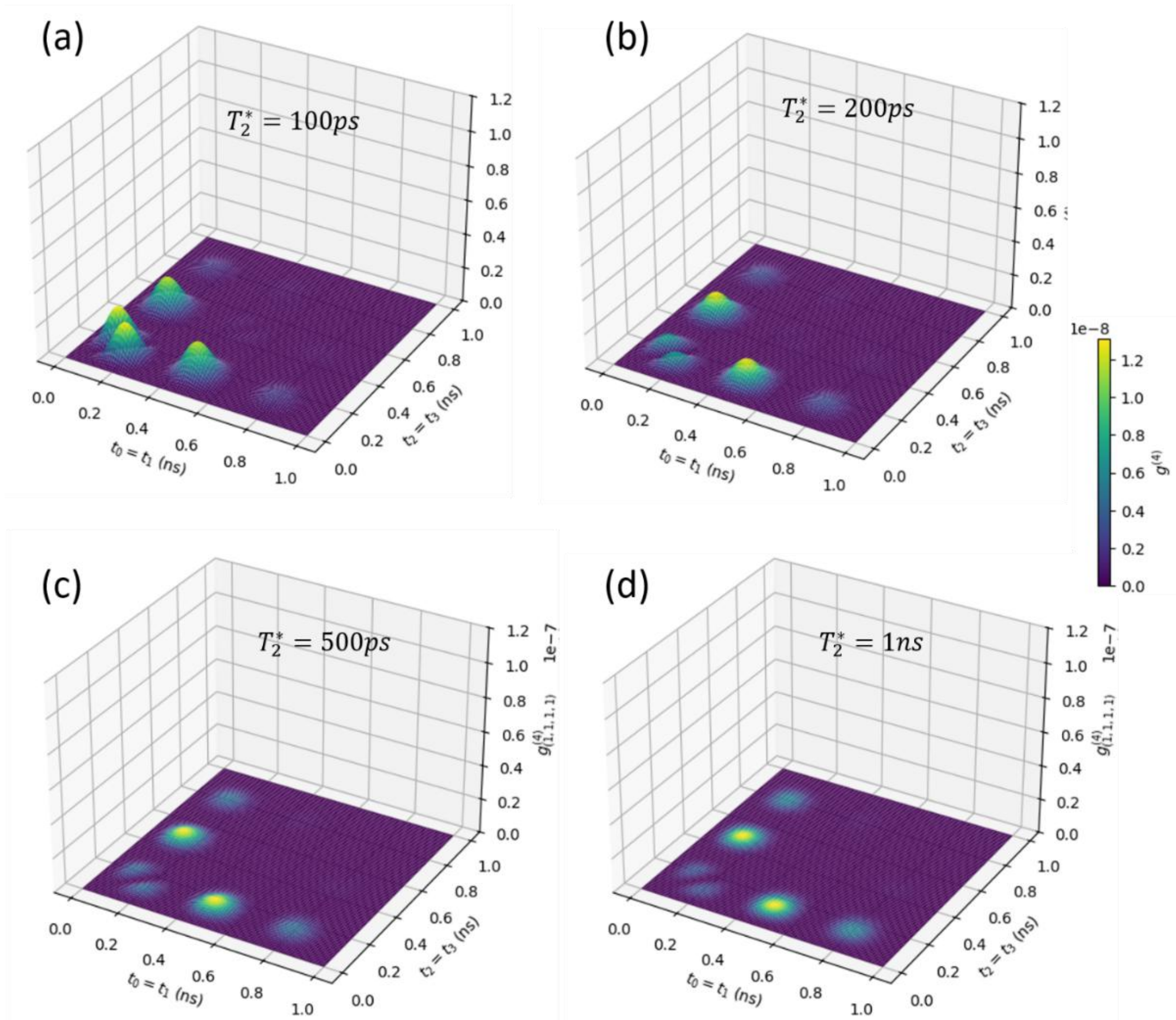

**Figure 7.** the 2D slice of the 4D $g^{(4)}_{0(1),1(1),2(1),3(1)}(t_0, t_1, t_2, t_3)$ function where we have set $t_0 = t_1$ and $t_2 = t_3$. The result is a 2D histogram with two time axes which is shown for various values of $T_2^*$ as examples, $T_2^* =$ (a) 100ps, (b) 200ps, (c) 500ps and (d) 1ns. Vertical axis units here are arbitrary, but comparable between the 4 plots.

**Time resolved g$^{(4)}$ spectra, --**

We define a 4-photon coincidence count measurement where detectors 0,1,2, and 3 trigger, respectively, at $t_{M0}$, $t_{M1}$, $t_{M2}$, $t_{M3}$. The resultant photon coincidence count spectra corresponding to the output state $|1,1,1,1\rangle$ can be denoted as $g^{(4)}_{0(1),1(1),2(1),3(1)}(t_{M0}, t_{M1}, t_{M2}, t_{M3})$. Using the approach outlined above we now compute the expected distribution of this $g^{(4)}_{0(1),1(1),2(1),3(1)}(t_{M0}, t_{M1}, t_{M2}, t_{M3})$ function *including the contribution from pure dephasing.* To visualize the temporal distribution of the histogram, we first take slice through the 4D plane of $g^{(4)}_{0(1),1(1),2(1),3(1)}(t_0, t_1, t_2, t_3)$ by setting $t_0 = t_1$ and $t_2 = t_3$. In this way we get 2D data $g^{(4)}_{0(1),1(1),2(1),3(1)}(t_0, t_0, t_2, t_2)$. We show the surface plot of this 2D slice for various values of $T_2^*$ in Fig.7. From these 2D plots, we extract the histogram with respect to $t_0 - t_2$. This is shown on Fig. 8 and it gives an equivalent sense compared to the $g^{(2)}(\tau)$ of the 2-photon HOM interference. The 2D plot of a 4D surface is, however, only a slice in a multidimensional plane and getting a complete and comprehensive intuitive understanding is difficult.

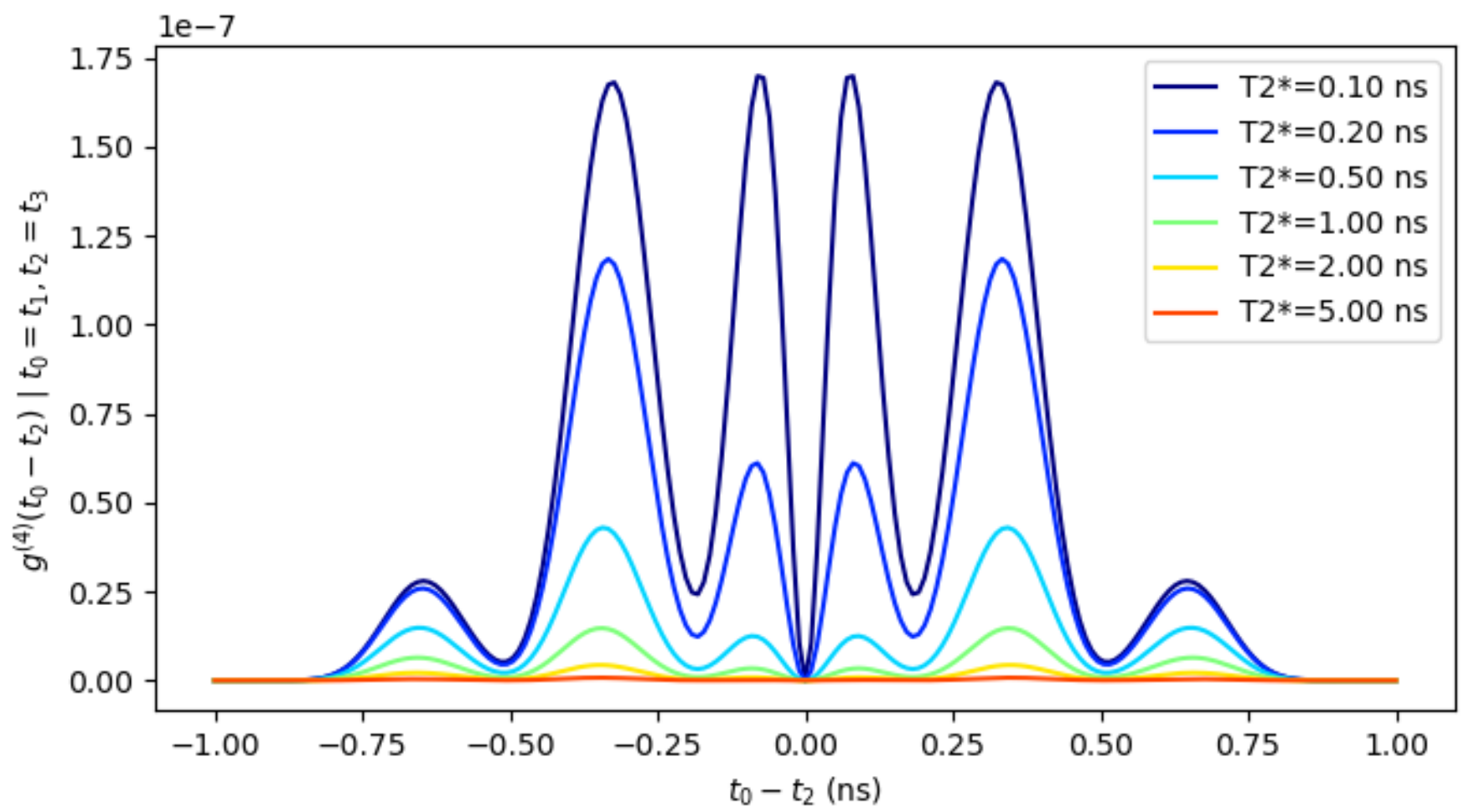

**Figure 8:** $g^{(4)}(t_0 - t_2)$ histogram given $t_0 = t_1$ and $t_2 = t_3$ for the (1,1,1,1) measured Fock basis of circuit shown in Fig. 10 for various values of the pure dephasing time constant $T_2^*$ of each single photon.

**IV: Discussions:**

In this work we have reported a theoretical and computational framework to estimate the output multi-dimensional wave packet produced by interference of a many photon wave packet in a linear optical circuit. We show explicit accounting for the multi-photon symmetrization by permuting the time axes of a multi-dimensional data representative of the many photon wave packets. The study is motivated by the recent development of the mesa-top single quantum dots-(MTSQDs)- a platform of on-chip QDs that can be fabricated in regular array, and provides pure, indistinguishable on-demand single photons resource states. As an example, we apply the analyses on a simple 4 photon 4 mode optical circuit with the wave packets mimicking the single photon states produced by the MTSQD SPS. We show that starting from a 4-photon product state as a resource state, there is probabilistic generation of 4 photon NOON state at the output. Using physically realistic parameters representing the single photon wave packets from MTSQD SPSs, we compute the effect of dephasing rate on the output probabilities of various Fock states of relevance, including the 4-photon NOON state. We also show how the dephasing rate affects the time-resolved 4 photon correlation count histogram.

The work paves the path to understanding effect of source-to-source variability between SPSs in an array and can provide guidance towards realization of large-scale quantum optical circuits based on on-demand many-photon resource states. Also, it can provide an opportunity to incorporate realistic responses of single photon and photon number resolving detectors in the measured correlation histogram. Our scheme disregards the symmetrization between photons in different modes which has no real impact on the measurement outcomes, and only accounts for the symmetrization of photons in the same mode. However, the dimension of the exponentially diverging permanent space still limits the number of photons/modes for which the effect of the photon wave packet on the output measurement outcome can be estimated. Further optimization to allow larger circuit modes, photons, and depth, can potentially be done using Finally, we note that at present the computational complexity is still limited by the exponentially diverging permanent space for the symmetry operation. Further optimization is needed to scale up the analysis to larger (>6 photon) circuits.

In summary, we provide a scheme to estimate the effect of the physical parameters of the photon resource states on the outcome in a linear optical circuit, e.g. in Boson sampling/ generation of entangled many-photon states via linear optics. We emphasize, that the Aaronson and Arkhipov scheme of Boson sampling (AABS) with deterministic Fock states as resource has been always known to provide access to more computational complexity compared to the squeezed Gaussian state-based Boson sampling (GBS). The current investigations in Boson sampling have been dominantly limited to GBS only because of lack of availibility of scalable ordered arrays of single photon sources so far. Recently the development of regular arrays of MTSQD single photon sources capable of producing deterministic/ on demand multi-photon resource states, has potentially made realization of large-scale qubit-based discreet variable Boson sampling i.e. AABS, possible. To this end, in this work we provide a way to directly connect to the wave packet shape and the dephasing characteristics of the single photons to the outcome of the linear optical quantum circuit and provide quantitative guidance to realization of scalable photonic QIP systems.

**Acknowledgement:**

The work is supported by US AFOSR grants FA9550-22-1-0376 and FA9550-25-1-0248

**References**

1. J. Zhang, S. Chattaraj, Q. Huang, L. Jordao, S. Lu, and A. Madhukar, On-chip scalable highly pure and indistinguishable single-photon sources in ordered arrays: Path to quantum optical circuits, Sci. Adv. **8**, eabn9252 (2022).

2. Q. Huang, S. Chattaraj, L. Jordao, J. Zhang, S. Lu, and A. Madhukar, Two-qubit gates using on-demand single-photons from ordered shape and size controlled large-volume superradiant quantum dots, Npj Quantum Inf **12**, 95 (2026).
3. C. M. Caves, Quantum-mechanical noise in an interferometer, Phys. Rev. D **23**, 1693 (1981).
4. D. J. Brod, E. F. Galvão, A. Crespi, R. Osellame, N. Spagnolo, and F. Sciarrinod, Photonic implementation of boson sampling: a review, Adv. Photon. 1, 1 (2019).
5. S. Aaronson and A. Arkhipov, *The Computational Complexity of Linear Optics*, in *Proceedings of the Forty-Third Annual ACM Symposium on Theory of Computing* (ACM, San Jose California USA, 2011), pp. 333–342.
6. M. AbuGhanem, Information processing at the speed of light, Front. Optoelectron. **17**, 33 (2024).
7. C. S. Hamilton, R. Kruse, L. Sansoni, S. Barkhofen, C. Silberhorn, and I. Jex, Gaussian Boson Sampling, Phys. Rev. Lett. **119**, 170501 (2017).
8. H. Wang et al., High-efficiency multiphoton boson sampling, Nature Photon **11**, 361 (2017).
9. H. Wang et al., Boson Sampling with 20 Input Photons and a 60-Mode Interferometer in a 1 0 14 -Dimensional Hilbert Space, Phys. Rev. Lett. **123**, 250503 (2019).
10. L. Jordao, S. Chattaraj, Q. Huang, S. Lu, J. Zhang, and A. Madhukar, Single photon superradiance enhanced light–matter interaction in spatially ordered shape and volume controlled single quantum dots: enabling on-chip photonic networks, Nanophotonics (2025).
11. J. Bylander, I. Robert-Philip, and I. Abram, Interference and correlation of two independent photons, Eur. Phys. J. D **22**, 295 (2003).
12. A. Thoma et al., Exploring Dephasing of a Solid-State Quantum Emitter via Time- and Temperature-Dependent Hong-Ou-Mandel Experiments, Phys. Rev. Lett. **116**, 033601 (2016).
13. J. P. Dowling, Quantum optical metrology – the lowdown on high-N00N states, Contemporary Physics **49**, 125 (2008).